\documentclass{cai}
\usepackage{graphicx}

\usepackage{amsmath}
\usepackage[toc,page]{appendix}
\usepackage{float}
\usepackage{rotating}
\usepackage{adjustbox}
\usepackage{hyperref}
\usepackage[colorinlistoftodos,textsize=scriptsize]{todonotes}
\usepackage{caption}
\usepackage{subcaption}
\usepackage{tikz}
\tikzstyle{detconst} = [rectangle, rounded corners, minimum width=2cm, minimum height=1cm, text centered, draw=black, fill=none, thick]
\tikzstyle{blank} = [rectangle, minimum width=0.1cm, minimum height=1cm, draw=none, fill=none]
\tikzstyle{arrow} = [thick,->,>=stealth]
\tikzstyle{arrowblank} = [thick,-]
\usepackage{ulem}

\usepackage{ifthen} 
\newboolean{uprightparticles}
\setboolean{uprightparticles}{false} 
\usepackage{xspace} 
\usepackage{upgreek}

\def\lhcb   {\mbox{LHCb}\xspace}

\def\cern {\mbox{CERN}\xspace}
\def\lhc    {\mbox{LHC}\xspace}

\def\velo   {VELO\xspace}
\def\rich   {RICH\xspace}
\def\richone {RICH1\xspace}
\def\richtwo {RICH2\xspace}

\def\ecal   {ECAL\xspace}
\def\hcal   {HCAL\xspace}
\def\MagUp {\mbox{\em Mag\kern -0.05em Up}\xspace}

\def\lone   {L0\xspace}
\def\hlt    {HLT\xspace}

\ifthenelse{\boolean{uprightparticles}}%
{

 \def\PDelta      {\ensuremath{\Delta}\xspace}                 
 \def\PXi         {\ensuremath{\Xi}\xspace}                 
 \def\PLambda     {\ensuremath{\Lambda}\xspace}                 
 \def\PSigma      {\ensuremath{\Sigma}\xspace}                 
 \def\POmega      {\ensuremath{\Omega}\xspace}                 
 \def\PUpsilon    {\ensuremath{\Upsilon}\xspace}

 \def\PB      {\ensuremath{\mathrm{B}}\xspace}                 
                  
 \def\PD      {\ensuremath{\mathrm{D}}\xspace}

 \def\PK      {\ensuremath{\mathrm{K}}\xspace}

 \def\Pb      {\ensuremath{\mathrm{b}}\xspace}                 
 \def\Pc      {\ensuremath{\mathrm{c}}\xspace}

 \def\Pi      {\ensuremath{\mathrm{i}}\xspace}

 \def\Ps      {\ensuremath{\mathrm{s}}\xspace}

 \def\thebaroffset{0.0em}
}
{

 \mathchardef\PDelta="7101
 \mathchardef\PXi="7104
 \mathchardef\PLambda="7103
 \mathchardef\PSigma="7106
 \mathchardef\POmega="710A
 \mathchardef\PUpsilon="7107
                  
 \def\PB      {\ensuremath{B}\xspace}                 
                  
 \def\PD      {\ensuremath{D}\xspace}

 \def\PK      {\ensuremath{K}\xspace}

 \def\Pb      {\ensuremath{b}\xspace}                 
 \def\Pc      {\ensuremath{c}\xspace}

 \def\Pi      {\ensuremath{i}\xspace}

 \def\Ps      {\ensuremath{s}\xspace}

 \def\thebaroffset{0.18em}
}
\newcommand{\offsetoverline}[2][\thebaroffset]{\kern #1\overline{\kern -#1 #2}}%

\makeatletter
\ifcase \@ptsize \relax
  \newcommand{\miniscule}{\@setfontsize\miniscule{4}{5}}
\or
  \newcommand{\miniscule}{\@setfontsize\miniscule{5}{6}}
\or
  \newcommand{\miniscule}{\@setfontsize\miniscule{5}{6}}
\fi
\makeatother

\DeclareRobustCommand{\optbar}[1]{\shortstack{{\miniscule (\rule[.5ex]{1.25em}{.18mm})}
  \\ [-.7ex] $#1$}}

\def\squark    {{\ensuremath{\Ps}}\xspace}

\def\cquark    {{\ensuremath{\Pc}}\xspace}

\def\bquark    {{\ensuremath{\Pb}}\xspace}

\def\KorKbar {\kern \thebaroffset\optbar{\kern -\thebaroffset \PK}{}\xspace}

\def\D       {{\ensuremath{\PD}}\xspace}

\def\DorDbar {\kern \thebaroffset\optbar{\kern -\thebaroffset \PD}\xspace}

\def\Dp      {{\ensuremath{\D^+}}\xspace}
\def\Dm      {{\ensuremath{\D^-}}\xspace}

\def\DpDm    {\ensuremath{\Dp {\kern -0.16em \Dm}}\xspace}

\def\B       {{\ensuremath{\PB}}\xspace}

\def\BorBbar {\kern \thebaroffset\optbar{\kern -\thebaroffset \PB}\xspace}

\def\Bd      {{\ensuremath{\B^0}}\xspace}

\def\BdorBdbar {\kern \thebaroffset\optbar{\kern -\thebaroffset \Bd}\xspace}

\def\Bs      {{\ensuremath{\B^0_\squark}}\xspace}

\def\BsorBsbar {\kern \thebaroffset\optbar{\kern -\thebaroffset \Bs}\xspace}

\def\Y#1S{\ensuremath{\PUpsilon{(#1S)}}\xspace}

\def\LorLbar     {\kern \thebaroffset\optbar{\kern -\thebaroffset \PLambda}\xspace}

\def\AT#1     {\ensuremath{A_{\mathrm{T}}^{#1}}\xspace}           

\def\C#1      {\ensuremath{\mathcal{C}_{#1}}\xspace}                       
\def\Cp#1     {\ensuremath{\mathcal{C}_{#1}^{'}}\xspace}                    
\def\Ceff#1   {\ensuremath{\mathcal{C}_{#1}^{\mathrm{(eff)}}}\xspace}        
\def\Cpeff#1  {\ensuremath{\mathcal{C}_{#1}^{'\mathrm{(eff)}}}\xspace}       
\def\Ope#1    {\ensuremath{\mathcal{O}_{#1}}\xspace}                       
\def\Opep#1   {\ensuremath{\mathcal{O}_{#1}^{'}}\xspace}                    

\newcommand{\aunit}[1]{\ensuremath{\text{\,#1}}}       

\newcommand{\tev}{\aunit{Te\kern -0.1em V}\xspace}
\newcommand{\gev}{\aunit{Ge\kern -0.1em V}\xspace}
\newcommand{\mev}{\aunit{Me\kern -0.1em V}\xspace}
\newcommand{\kev}{\aunit{ke\kern -0.1em V}\xspace}
\newcommand{\ev}{\aunit{e\kern -0.1em V}\xspace}
 
\newcommand{\mevc}{\ensuremath{\aunit{Me\kern -0.1em V\!/}c}\xspace}
\newcommand{\gevc}{\ensuremath{\aunit{Ge\kern -0.1em V\!/}c}\xspace}
\newcommand{\mevcc}{\ensuremath{\aunit{Me\kern -0.1em V\!/}c^2}\xspace}
\newcommand{\gevcc}{\ensuremath{\aunit{Ge\kern -0.1em V\!/}c^2}\xspace}

\def\mhz  {\ensuremath{\aunit{MHz}}\xspace}

\def\gsim{{~\raise.15em\hbox{$>$}\kern-.85em
          \lower.35em\hbox{$\sim$}~}\xspace}
\def\lsim{{~\raise.15em\hbox{$<$}\kern-.85em
          \lower.35em\hbox{$\sim$}~}\xspace}

\def\boole      {\mbox{\textsc{Boole}}\xspace}
\def\brunel     {\mbox{\textsc{Brunel}}\xspace}

\def\evtgen     {\mbox{\textsc{EvtGen}}\xspace}

\def\gaudi      {\mbox{\textsc{Gaudi}}\xspace}
\def\gauss      {\mbox{\textsc{Gauss}}\xspace}
\def\geant      {\mbox{\textsc{Geant4}}\xspace}

\def\moore      {\mbox{\textsc{Moore}}\xspace}

\def\pythia     {\mbox{\textsc{Pythia}}\xspace}
\def\pythiaeight     {\mbox{\textsc{Pythia8}}\xspace}

\def\ddforhep     {\mbox{\textsc{DD4hep}}\xspace}
\def\gaussino     {\mbox{\textsc{Gaussino}}\xspace}
\def\gog    {\mbox{\textsc{Gauss-on-Gaussino}}\xspace}

\def\externaldetector {\mbox{\textsc{ExternalDetector}}\xspace}
\def\mccollector {\mbox{\textsc{MCCollector}}\xspace}
\def\parallelgeometry {\mbox{\textsc{ParallelGeometry}}\xspace}
\def\fastsimulation {\mbox{\textsc{FastSimulation}}\xspace}

\def\cpp        {\mbox{\textsc{C\raisebox{0.1em}{{\footnotesize{++}}}}}\xspace}

\def\gbytes     {\aunit{GB}\xspace}

\def\tell1  {TELL1\xspace}
\def\ukl1   {UKL1\xspace}

\def\fifo   {FIFO\xspace}

\usepackage[symbol]{footmisc}

\newcommand\blfootnote[2]{%
  \begingroup
  \renewcommand\thefootnote{}\footnote{#2}%
  \addtocounter{footnote}{-1}%
  \endgroup
}

\begin{document}
\label{firstpage}

\title[New simulation software technologies at the LHCb Experiment at CERN]
      {New simulation software technologies \\ at the LHCb Experiment at CERN}

\author[M. Mazurek, G. Corti, D. M\"{u}ller]
       {Micha\l{} \surname{Mazurek}$^{1, 2}$, Gloria \surname{Corti}$^1$, Dominik \surname{M\"{u}ller}$^1$}
\affiliation{$^1$CERN, European Organization for Nuclear Reasearch\\
Esplanade des Particules 1, 1211 Meyrin, Switzerland \\ $^2$NCBJ, National Centre for Nuclear Research\\
Andrzeja Sołtana 7, 05-400 Otwock, Poland}
\email{michal.mazurek@cern.ch, gloria.corti@cern.ch}

%
%

\noreceived{} \nocommunicated{}

\maketitle

\begin{abstract}
The \lhcb experiment at the Large Hadron Collider (\lhc) at \cern has successfully performed a large number of physics measurements during Runs 1 and 2 of the \lhc. Monte Carlo simulation is key to the interpretation of these and future measurements. The \lhcb experiment is currently undergoing a major detector upgrade for Run 3 of the \lhc to process events with five times higher luminosity. New simulation software technologies have to be introduced to produce simulated data samples of sufficient size within the computing resources allocated for the next few years. Therefore, the \lhcb collaboration is currently preparing an upgraded version of its \gauss simulation framework. The new version provides the \lhcb specific functionality while its generic simulation infrastructure has been encapsulated in an experiment independent framework, \gaussino. The latter combines the \gaudi core software framework and the \geant~simulation toolkit and fully exploits their multi-threading capabilities. A prototype of a fast~simulation~interface to the simulation toolkit is being developed as the latest addition to \gaussino to provide an extensive palette of fast simulation models, including new deep learning-based options.
\end{abstract}

\begin{keywords}
LHCb, CERN, Monte Carlo simulation, fast simulations,\\
multi-threading
\end{keywords}

\begin{mathclass}
68-04
\end{mathclass}

\begin{center}
    Submitted to Computing and Informatics
\end{center}

\section{Introduction}

The \lhcb detector is a  single-arm spectrometer with a forward angular coverage, designed for the study of particles containing \bquark or \cquark quarks. It is located at the Large Hadron Collider (\lhc) at the European Organization for Nuclear Research (\cern). During the LHC Run 1 and Run 2 the \lhcb experiment has successfully performed a large number of measurements in heavy flavour physics, however, the precision of its measurements is becoming more and more limited by the statistics of the simulated samples. In addition, the LHCb experiment is currently installing a major upgrade for the data taking of Run 3 in order to be able to process events with a 5 times higher luminosity. The \lhcb experiment is also planning a further upgrade with another increase in luminosity by a factor of 5 to 10. Therefore, higher capacity in data storage and computing power is needed in order to prepare the experiment for the changes in the future Runs. Moreover, the whole software is being adapted to work in a multi-threaded environment in order to efficiently use the available computing resources. As an example, the experiment will rely on a real-time reconstruction trigger implemented on GPUs. In the last years, the \lhcb \gauss simulation framework has been evolving to provide a palette of fast simulations options and to use the new functional \gaudi framework and \geant multi-threading  technologies. In this paper, recent developments in the simulation software are discussed. A brief description of the \lhcb detector and its upgrade are presented in Section~\ref{section:detector}. Details about the \lhcb simulation software and the new multi-threading approach can be found in Section~\ref{section:software}. Finally, a description of the fast simulation interface in \gog with an example on how to use the libraries needed to prepare a training dataset is presented in Section~\ref{section:fastsim}.

\section{LHCb Detector}
\label{section:detector}
The \lhcb detector~\cite{LHCb-DP-2008-001,LHCb-DP-2014-002} covers the \mbox{pseudorapidity} range $2<\eta <5$. It includes a high-precision tracking system consisting of a silicon-strip vertex detector (VErtex LOcator, \velo) surrounding the $pp$ interaction region, a large-area silicon-strip detector (TT) located upstream of a dipole magnet with a bending power of about $4{\mathrm{\,Tm}}$, and three stations of silicon-strip detectors and straw drift tubes placed downstream of the magnet. The tracking system (T1-T3) provides the momentum measurement of charged particles. Different types of charged hadrons are distinguished using information from two ring-imaging Cherenkov detectors, located upstream and downstream of the magnet (\richone and \richtwo). Photons, electrons and hadrons are identified by the \lhcb calorimeter system~\cite{LHCb-DP-2020-001}.  An Electromagnetic CALorimeter (\ecal) and a Hadronic CALorimeter (\hcal)  determine the amount of energy deposited by the particles of electromagnetic or hadronic nature, respectively. Some minimum ionizing particles may also be registered by the calorimeters. Muons are identified by a system composed of alternating layers of iron and multiwire proportional chambers (M1-M5). In order to reduce the number of events collected and maximizing those of interest at the same time, an online event selection is performed via a combination of hardware and software triggers. The \lone trigger (hardware stage) operates on the information coming from the calorimeter and muon systems at a rate of 40 \mhz, followed by a High Level Trigger (\hlt, software stage), which applies a full event reconstruction.



The \lhcb collaboration is currently installing a  major upgrade that will allow the experiment to significantly increase the amount of data collected, by operating with a luminosity of up to $2 \times 10^{33}\;\mathrm{cm}^{-2}\mathrm{s}^{-1}$, 5 times higher than that of Run 1 and Run 2 and writing data at a 30 \mhz rate. In order to cope with the higher readout rates, the front-end electronics of all the sub-detectors will be replaced. Moreover, most of the \lhcb sub-detectors will be replaced and only the heaviest ones, namely the calorimeters and muon stations, will remain mostly in place. In Run 1 and Run 2 the event output rate was limited to 1 \mhz and through a hardware-based \lone trigger. This has been removed for Run 3 and LHCb is introducing a software-only trigger working at the 30 \mhz bunch crossing rate. Processing and the selection of events at this very high rate have a major impact on software and computing systems.

\section {Simulation software at \lhcb}
\label{section:software}

\begin{figure}
    \centering
    \includegraphics[width=\textwidth]{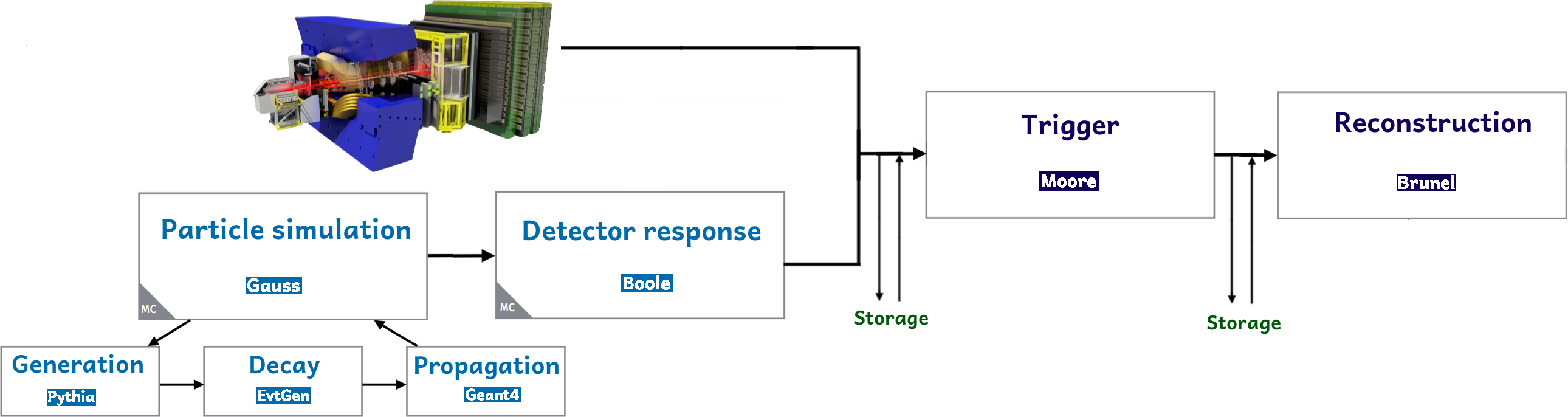}
    \caption{Simplified view on the LHCb data flow and data processing applications as used in Run 1 and Run 2.}
    \label{fig:lhcb_data_flow}
\end{figure}

In \lhcb, data collected by the detector is processed using a set of custom applications based on the \gaudi \cite{Gaudi, Gaudi2} core software framework. A simplified view on the \lhcb data processing applications, as used in Run 1 and Run 2, is shown in Figure~\ref{fig:lhcb_data_flow}. \lhcb applications identically process events either collected by the detector itself or events produced by the simulation software. Simulated events are first handled by \gauss that performs the event generation and tracking of particles through the detector material. The \boole application mimics the specific sub-detector technologies and electronics response, providing the same digital output of the data acquisition system. The \moore application then emulates the \lone trigger response for the simulated data and provides the software-based \hlt trigger. More complex objects are produced by the offline reconstruction in \brunel. Each application reads and writes processed data to storage.

\subsection {\gauss simulation software}

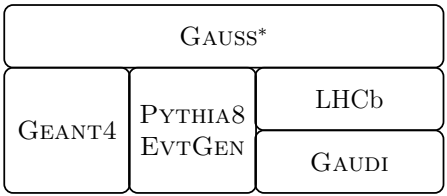
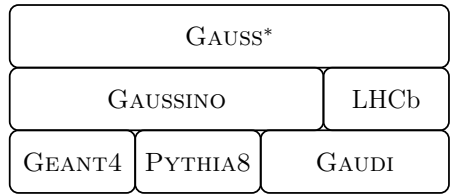
\begin{figure}
\centering
\begin{subfigure}[h]{0.48\textwidth}
\begin{tikzpicture}[node distance=0cm]
\node (g4) [detconst, minimum height=2cm] {\geant};
\node (pt) [detconst, right of=g4, xshift=2cm, minimum height=2cm,text width=1.75cm] {\pythiaeight \evtgen};
\node (gaudi) [detconst, right of=pt, yshift=-.5cm, xshift=2.5cm, minimum width=3cm] {\gaudi};
\node (lhcb) [detconst, above of=gaudi, yshift=1cm, minimum width=3cm] {\lhcb};
\node (gauss) [detconst, above of=g4, yshift=1.5cm, xshift=2.5cm, minimum width=7cm] {\gauss\footnotemark};
\end{tikzpicture}
\caption{\gauss Run 1 and Run 2 framework.}
\label{fig:software_stack_run1}
\end{subfigure}
\hfill
\begin{subfigure}[h]{0.48\textwidth}
\begin{tikzpicture}[node distance=0cm]
\node (g4) [detconst] {\geant};
\node (pt) [detconst, right of=g4, xshift=2cm] {\pythiaeight};
\node (gaudi) [detconst, right of=pt, xshift=2.5cm, minimum width=3cm] {\gaudi};
\node (lhcb) [detconst, above of=gaudi, yshift=1cm, xshift=.5cm] {\lhcb};
\node (gaussino) [detconst, above of=g4, yshift=1cm, xshift=1.5cm, minimum width=5cm] {\gaussino};
\node (gauss) [detconst, above of=g4, yshift=2cm, xshift=2.5cm, minimum width=7cm] {\gauss\footnotemark[\value{footnote}]};
\end{tikzpicture}
\caption{\gog framework}
\label{fig:software_stack_upgrade}
\end{subfigure}
\caption{Dependencies~\cite{FastSimInterface} in the simulation software stack before and after upgrade.}
\end{figure}

The \lhcb simulation framework \gauss~\cite{Gauss} is built, as all the other \lhcb software applications, using the general \gaudi data-processing framework. \gaudi helps in the configuration of algorithms and tools in the application. It also controls the data flow and the event loop. Dependencies in the simulation framework are illustrated in Figure \ref{fig:software_stack_run1}. One of the two main tasks of the \gauss framework is to control the generation of collisions, in most cases $pp$ with \pythia~\cite{Sjostrand:2006za, Sjostrand:2007gs}, where a specific \lhcb configuration~\cite{LHCb-PROC-2010-056} is used. \gauss then makes use of \evtgen~\cite{Lange:2001uf} in order to model the decays of unstable particles. The other main task of \gauss is the propagation of the generated particles through the experimental apparatus and the simulation of physics processes occurring within the sub-detectors using the \geant toolkit~\cite{Allison:2006ve, Agostinelli:2002hh, LHCb-PROC-2011-006}. The  information to be able to mimic the response of given sub-detectors when a particle intersects them (MC hits) and to understand trigger and reconstruction performance (MC truth) are written to a file for further processing.

\subsection{\lhcb software changes for Run 3}

 As mentioned in the first section, providing a high-throughput software relying on a better utilization of modern computing architectures is crucial in order to cope with the higher event output rate. A typical \lhcb application needs around 2 to 4 \gbytes RAM on average \cite{GaudiUpgrade}. Therefore, a multi-process approach, where each instance of the application is run on a separate core, is not optimal from the memory consumption point of view. As a result, the \lhcb core software framework, \gaudi, has been re-engineered to work in a multi-threading mode in order to benefit from the inter-event parallelism as much as possible. A lot of effort has been put in order to guarantee thread-safety, as many objects in the application are now shared between the events. This has been achieved mostly by introducing a functional paradigm, in which a set of algorithms operate on data by either producing, transforming or consuming it, and ensuring that the internal state is unchanged. The implementation of a new scheduler \cite{LHCb-TALK-2018-343} and the optimization of the software by utilizing vector processing units, introducing new \cpp features and rethinking the data models, were the other key components in the new \gaudi framework.

\footnotetext{Additional LHCb-specific configuration for \geant, \pythia and \evtgen is not shown in the dependency graph.}

\gauss, the \lhcb simulation software, like all the other \lhcb applications, depends on the changes introduced in \gaudi. Moreover, \gauss has to adapt to the changes introduced by the new approach in the detector description. \ddforhep \cite{dd4hep} is a detector description toolkit that provides experiment-independent libraries for conditions management and visualization of the geometry.  On the other hand, \gauss still has to provide support for the older geometries when producing simulated samples for Run~1 and Run~2 physics analyses. Additional challenge arises from the fact that \gauss delegates tasks to external libraries used in the framework (e.g. \geant) that are already working in their own multi-threading realms.

\subsection{Gauss-on-Gaussino}

In order to address the challenges mentioned above, the \lhcb simulation team decided to move all the \lhcb-independent components from the simulation software and place it in a separate project, called \gaussino \cite{Gaussino, Gaussino2, Gaussino3}, as a core simulation framework, on which all the new versions of \gauss will be built. It is also possible to run \gaussino as a standalone application. Dependencies in the new \lhcb version of the framework, called here \gog for clarity, are illustrated in Figure \ref{fig:software_stack_upgrade}. \gaussino follows the \gaudi's inter-event-based parallelism of the event loop, in which algorithms are scheduled in a way that guarantees thread-safety. \gaussino communicates with \geant objects by creating corresponding factories that act as  \gaudi tools. In the event loop, \gaudi places the generated event on the top of a \fifo queue. \geant worker threads then take the event from the queue and perform the simulation following their own multi-threading scheme. The multi-threaded approach in \gog has already made it possible to simulate more events in time by limiting the memory consumption of each event as shown in an earlier paper~\cite{Gaussino3}. Nevertheless, this will not be enough to meet the requirements imposed by the upgrade of the detector, which will be discussed in the following section.

\section{Fast simulation interface in Gauss-on-Gaussino}
\label{section:fastsim}

\begin{figure}
    \centering
    \includegraphics[width=.83\textwidth]{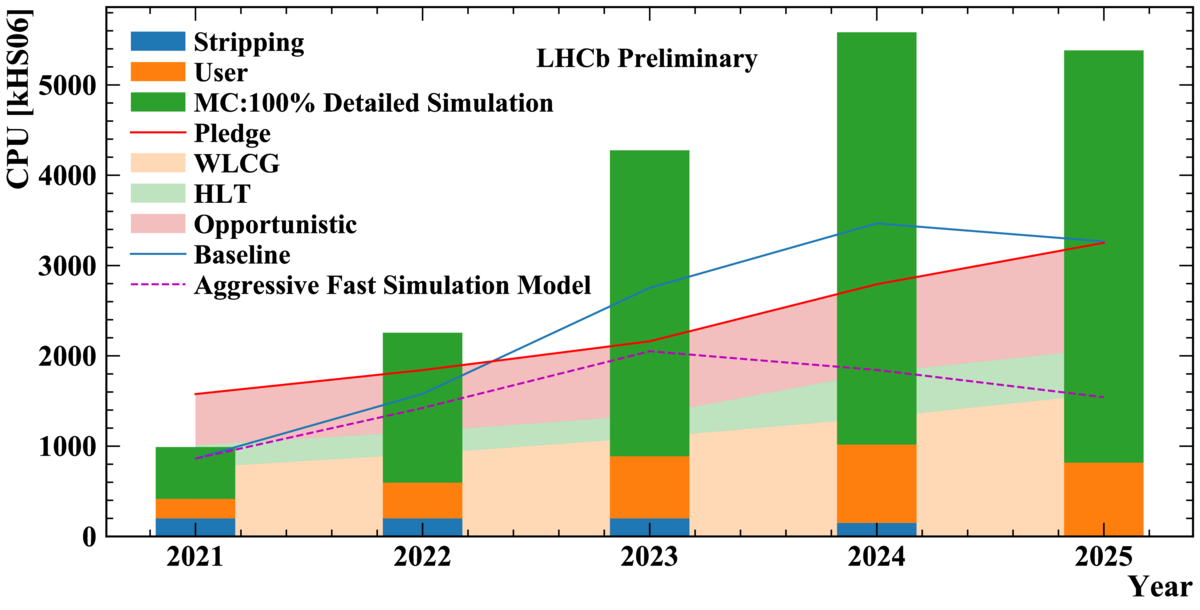}
    \caption{Projection of the computing resources available to \lhcb \cite{LHCB-FIGURE-2019-018}. Monte Carlo simulations will quickly exceed in the upcoming years the resources allocated to \lhcb by the distributed computing system provided that no action is taken in order to fit in the requirements (e.g. fast simulations). The CPU needs for producing 100\% of the required samples with the detailed simulation is shown by the green bars. The CPU needs in two scenarios with different fractions of events produced with detailed, fast and ultra-fast simulations are shown by the blue and dashed purple lines respectively.}
    \label{fig:computingresources}
\end{figure}

Producing necessary simulated samples for physics analyses with \gauss consumed around 80\% \cite{LHCB-FIGURE-2019-018, Whitehead, LHCb-TDR-018} of all the distributed computing resources available to \lhcb during Run 2. An additional few percent were used to process the simulated samples through the subsequent applications. The increase in the number of events, caused by the upcoming upgrade, will require to simulate even more events. If all samples would be produced with a detailed simulation, the foreseen computing resources allocated to \lhcb will be quickly exceeded (Figure \ref{fig:computingresources}). Scenarios with different
fractions of the samples produced with detailed simulation, fast, and ultra-fast models would allow to fulfill the needs within the resource forecast. 

\begin{figure}
\begin{subfigure}[h]{\textwidth}
    \centering
    \includegraphics[width=.85\textwidth]{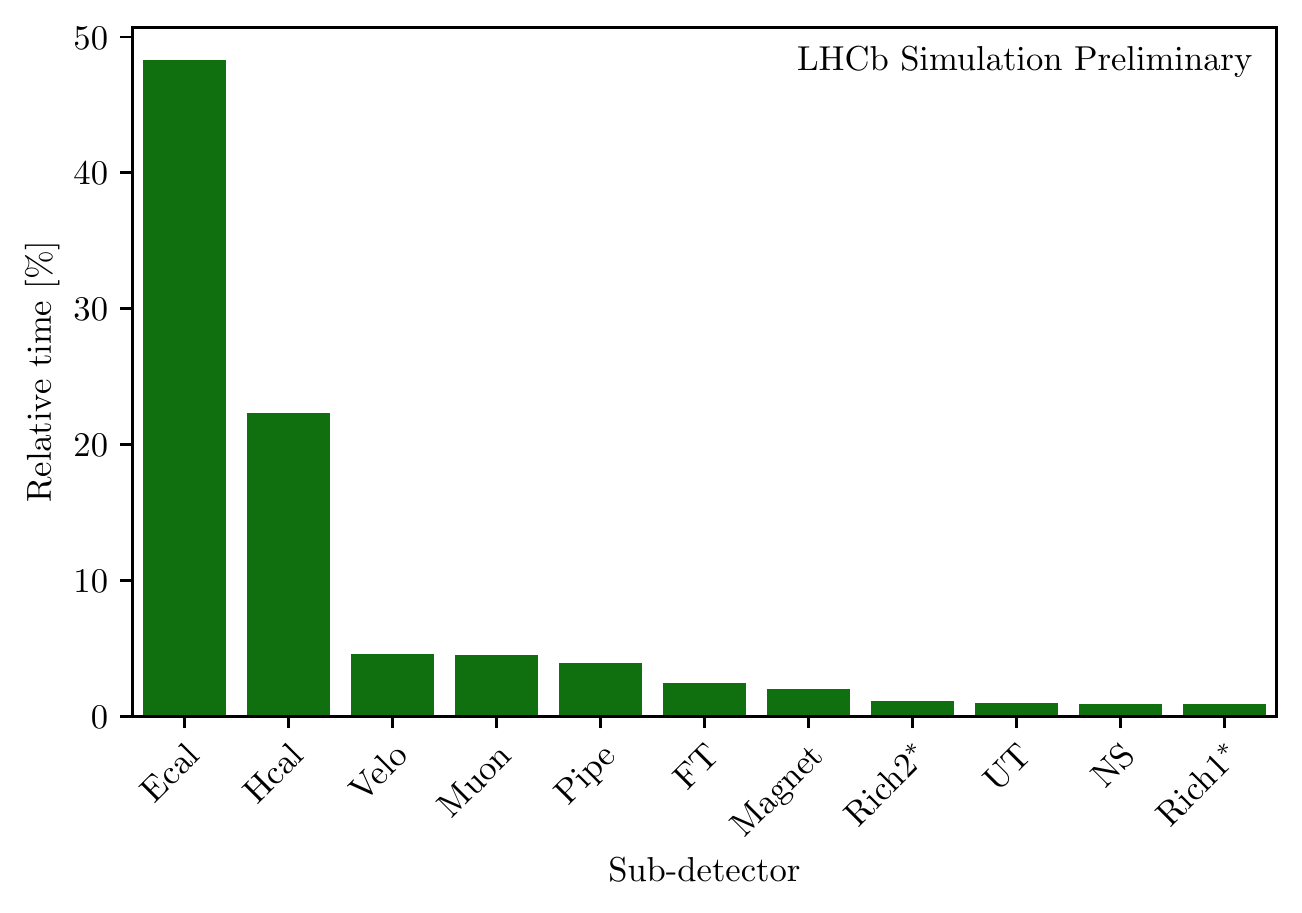}
    \caption{Current \gauss framework with \geant v10.6.}
    \label{fig:current_gauss_minbias_upgrade_timing_per_detector}
\end{subfigure}
\blfootnote{*}{$^*$ Simulation of the optical photons in \richone and \richtwo was turned off.}
\begin{subfigure}[h]{\textwidth}
    \centering
    \includegraphics[width=.85\textwidth]{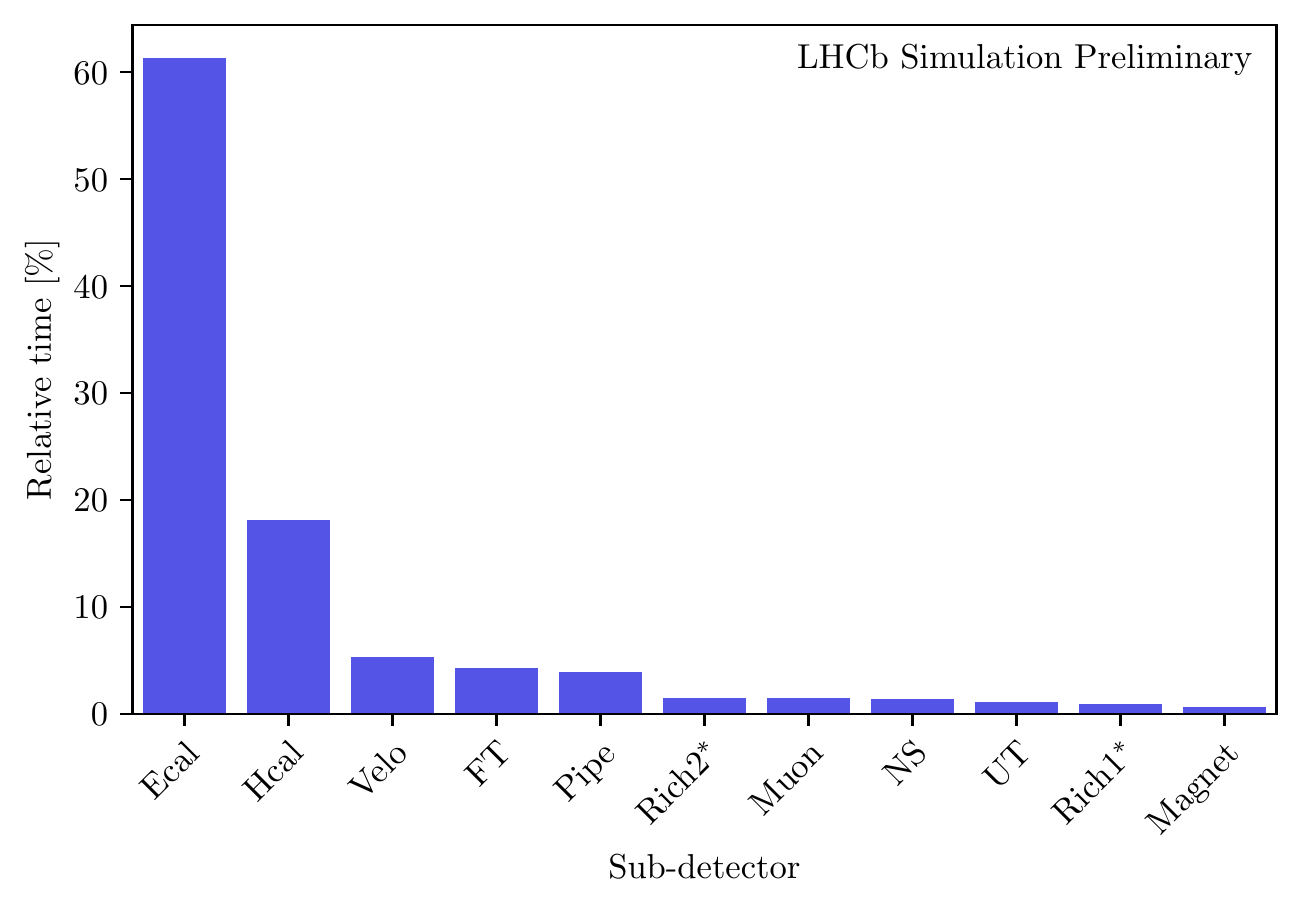}
    \caption{New multi-threaded (1 thread) \gog framework with \geant v10.7.}
    \label{fig:new_gauss_minbias_upgrade_timing_per_detector}
\end{subfigure}
    \caption{Relative time~\cite{FastSimInterface} spent in each sub-detector when simulating 100 minimum bias events using different versions of the framework with the beam conditions as foreseen in the Run 3 data-taking period and the upgrade geometry. The time spent in the calorimeters is very similar in the current and new versions of the framework as expected. Simulation of the optical photons in \richone and \richtwo was turned off in the current version of \gauss as it is not yet available in \gog, for easier comparison.}
\end{figure}

A lot of efforts have been made by the \gauss developers in order to reduce the CPU time spent in the simulation. 
In \gauss, propagating particles through matter dominates the time spent by the application. A finer analysis of the relative time spent in each sub-detector of the upgrade geometry is presented in Figure \ref{fig:current_gauss_minbias_upgrade_timing_per_detector} for the current version of the framework. As seen in the results, simulation of the particle showers in the calorimeters is especially demanding. Consistent results are obtained for \gog as expected and shown in Figure \ref{fig:new_gauss_minbias_upgrade_timing_per_detector}. Additional information about the time spent by each particle is given in Figure \ref{fig:reltive_time_upgrade} and Figure \ref{fig:reltive_time_2016} in the Appendix.  

A campaign with the goal to introduce a palette of fast simulation models to complement the detailed \geant-based simulation has been launched.
ReDecay~\cite{LHCb-DP-2018-004} is a technique, in which the underlying $pp$ interaction is reused in the simulation of the detector multiple times, with an independently generated signal decay for each event. Lamarr \cite{LamarrPerformance} is an in-house, ultra-fast parametrization framework that extends up to the reconstruction level and provides high-level reconstruction objects in the output. When it comes to the calorimeters, a fast simulation model of \ecal based on a point library \cite{PointLib} is currently being implemented.  There are already a few proposals of fast simulation models based on generative adversarial networks \cite{GANs} for the calorimeters and \rich detectors, however, a special interface is needed in \gaussino to exploit them via the fast simulation mechanisms available in \geant. Recent work to support the implementation of these last fast simulations in \gaussino is described in the rest of this section.

\subsection{Implementation of the interface}

\begin{figure}
    \centering
    \begin{tikzpicture}[node distance=0cm,scale=.85,every node/.style={scale=.85}]
        \node (gaussino) [detconst, minimum width=6cm, minimum height=10cm, loosely dashed] {};
        \node (gaussinotext) [blank, xshift=1.5cm, yshift=-.25cm] at (gaussino.north west) {Gaussino};
        \node (fastsim) [detconst, yshift=0.75cm, above of=gaussino, minimum width=5.5cm, minimum height=7.25cm, loosely dotted] {};
        \node (fastsimtext) [blank, xshift=2cm, yshift=-.25cm] at (fastsim.north west) {Fast Simulation};
        \node (physicsfactory) [detconst, above of=fastsim, minimum width=5cm, yshift=2cm] {PhysicsFactory};
        \node (regionfactory) [detconst, above of=fastsim, minimum width=5cm, yshift=-.25cm] {RegionFactory};
        \node (modelfactory) [detconst, above of=fastsim, minimum width=5cm, yshift=-2.5cm] {ModelFactory};
        \node (gaussinorest) [blank, yshift=-4cm, above of=gaussino, minimum width=5.5cm, minimum height=2cm] {};
        \node (sensdetfactory) [detconst, above of=gaussinorest, minimum width=5cm] {DetectorFactory};

        \node (g4) [detconst, minimum width=6cm, minimum height=10cm, xshift=8cm, loosely dashed] {};
        \node (g4fastsim) [blank, yshift=.75cm, above of=g4, minimum width=5.5cm, minimum height=7.25cm] {};
        \node (g4text) [blank, xshift=1.5cm, yshift=-.25cm] at (g4.north west) {Geant4};
        \node (g4physics) [detconst, above of=g4fastsim, minimum width=5cm, yshift=2cm] {G4FastSimulationPhysics};
        \node (g4region) [detconst, above of=g4fastsim, minimum width=5cm, yshift=-.25cm] {G4Region};
        \node (g4model) [detconst, above of=g4fastsim, minimum width=5cm, yshift=-2.5cm] {G4VFastSimulationModel};
        \node (g4rest) [blank, yshift=-4cm, above of=g4, minimum width=5.5cm, minimum height=2cm] {};
        \node (g4sensdet) [detconst, above of=g4rest, minimum width=5cm] {G4VSensitiveDetector};

        \draw [arrow] (physicsfactory) -- node [text width=1.25cm, align=left, font=\scriptsize\linespread{0.8}\selectfont, midway,above] {construct()}(g4physics);
        \draw [arrow] (regionfactory) -- node [text width=1.25cm, align=left, font=\scriptsize\linespread{0.8}\selectfont, midway,above] {construct()}(g4region);
        \draw [arrow] (modelfactory) -- node [text width=1.25cm, align=left, font=\scriptsize\linespread{0.8}\selectfont, midway,above] {construct()}(g4model);
        \draw [arrow] (sensdetfactory) -- node [text width=1.25cm, align=left, font=\scriptsize\linespread{0.8}\selectfont, midway,above] {construct()}(g4sensdet);

        \node (gaussinolefttop)    [blank, xshift=-1cm] at (gaussino.north west) {};
        \node (gaussinoleftbottom) [blank, xshift=-1cm] at (gaussino.south west) {};
        \node (g4righttop)    [blank, xshift=1cm] at (g4.north east) {};
        \node (g4rightbottom) [blank, xshift=1cm] at (g4.south east) {};
        
        \draw [arrow] (gaussinolefttop) -- node [text width=10cm, align=center, font=\large\linespread{0.8}\selectfont, midway,sloped,anchor=center,below] {Sensitive Detector Construction}(gaussinoleftbottom);
        
        \draw [arrow] (g4righttop) -- node [text width=10cm, align=center, font=\large\linespread{0.8}\selectfont, midway,sloped,anchor=center,above] {Hit Extraction}(g4rightbottom);
    \end{tikzpicture}
    \caption{A simplified model~\cite{FastSimInterface} of the \fastsimulation interface with a set of dedicated factories that construct the corresponding \geant objects.}
    \label{fig:fast_simulation_flow}
\end{figure}
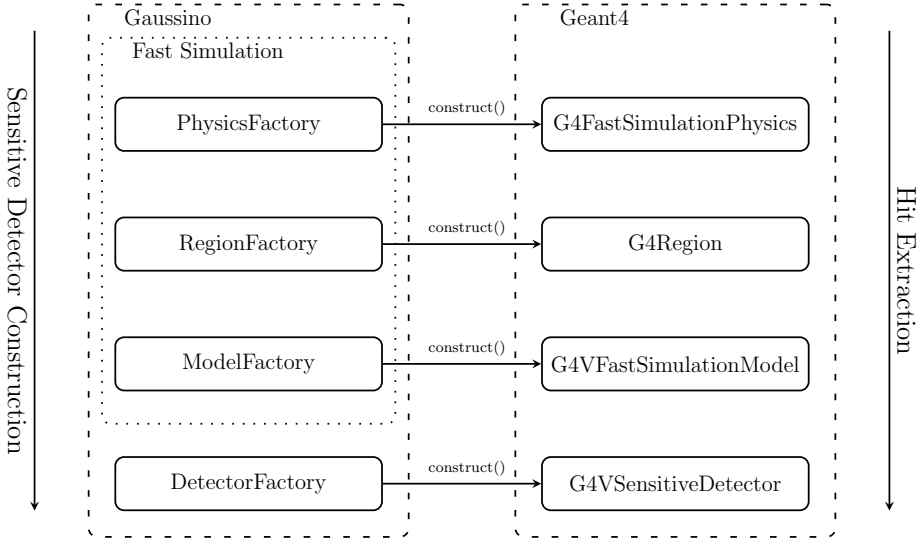

\gaussino provides a generic \fastsimulation interface to \geant objects in order to minimize the work spent in the future on implementing fast simulation models, and also to guarantee the integrity of the simulated data. Following the convention already present in \gaussino, the \fastsimulation interface consists of factories  ensuring \geant objects are configured properly and at the right moment when running the application. A set of the most important factories and their \geant counterparts are presented in Figure \ref{fig:fast_simulation_flow}. Most of the work necessary to implement a fast simulation model itself can be pushed to the configuration in python files. The most optimistic scenario is that the developer will only have to implement a \texttt{G4VFastSimulationModel::DoIt()} callback method in \cpp that is the key component of the whole interface and actually describes the whole process of generating fast hits.

\overfullrule=0pt

Two, purely abstract, fast simulation models are introduced in \gaussino in order to measure the performance of the interface, and to mark off a lower bound for all further fast simulation models. An \texttt{ImmediateDeposit} model generates one hit per particle that intercepts the region where the \texttt{ImmediateDeposit} model is active and deposits all of its energy in that hit. \texttt{ImmediateDeposit} gives useful information about the timing needed for the infrastructure itself to call the fast simulation methods. The \texttt{ShowerDeposit} model works in a similar manner, but it splits the energy of a particle into a selected number of hits, and generates them randomly around the position where the particle intercepted the region. \texttt{ShowerDeposit} provides the minimum amount of time needed to generate a specific number of hits with no additional calculations. 

\begin{figure}
    \centering
    \includegraphics[width=\textwidth]{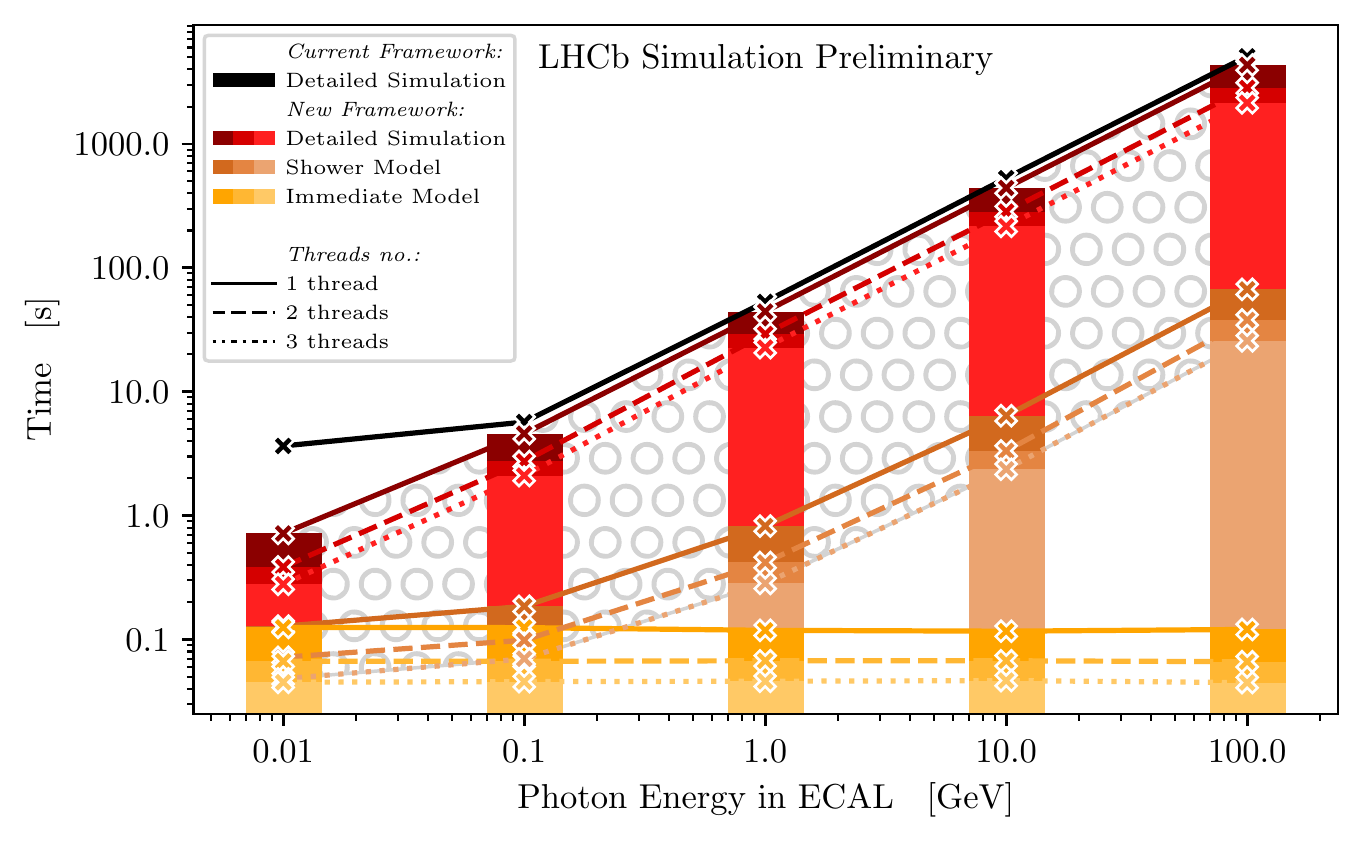}
    \caption{Comparison of the time~\cite{FastSimInterface} spent by different fast simulation models (\texttt{ImmediateDeposit} and \texttt{ShowerDeposit)} and a detailed simulation with \geant in the electromagnetic calorimeter. In each of the models tested, a particle gun generates a grid of evenly-spaced photons of a particular energy. For the detailed simulation the time of the current version of \gauss is also given as reference.}
    \label{fig:timing}
\end{figure}

A comparison between different fast simulation models tested with a particle gun that creates a grid of 3328 evenly-spaced photons originating at the \lhcb interaction point, is presented in Figure \ref{fig:timing}. Time spent in the \texttt{ImmediateDeposit} model is comparable across different photon energies and works as expected. On the other hand, the time spent in the \texttt{ShowerDeposit} model increases with the energy of photons. Naturally, this is caused by a larger number of hits generated in each shower. In the worst case, a 100 \gev photon generates 21558 hits on average in the calorimeter. Only around 25 seconds are needed for the fast simulation infrastructure in \gaussino with 3 threads to simulate 3328 x 21558 hits. In a detailed simulation, the time needed to simulate the same number of hits rises to 2145 seconds. The results prove that the infrastructure, used by the \fastsimulation interface, provides the possibility to significantly improve the time spent by the simulation software in the detector, provided  the  fast simulation model gives a similar level of precision in physics.

\subsection{Fast simulation training datasets}

 Many of the advanced, fast simulation models require prior tuning or training on some input data in order to provide valid results. When implementing a fast simulation model, a developer specifies a region that does not necessarily have to coincide with the \lhcb sub-detectors boundaries. The information required to train these models is not always available in an output file, as \gauss stores only the minimum amount of information required for physics studies. Therefore, the developer should be given the possibility to turn off any unneeded optimization features and to gather information at any given place in the detector.

\begin{figure}
    \centering
    \includegraphics[width=.6\textwidth]{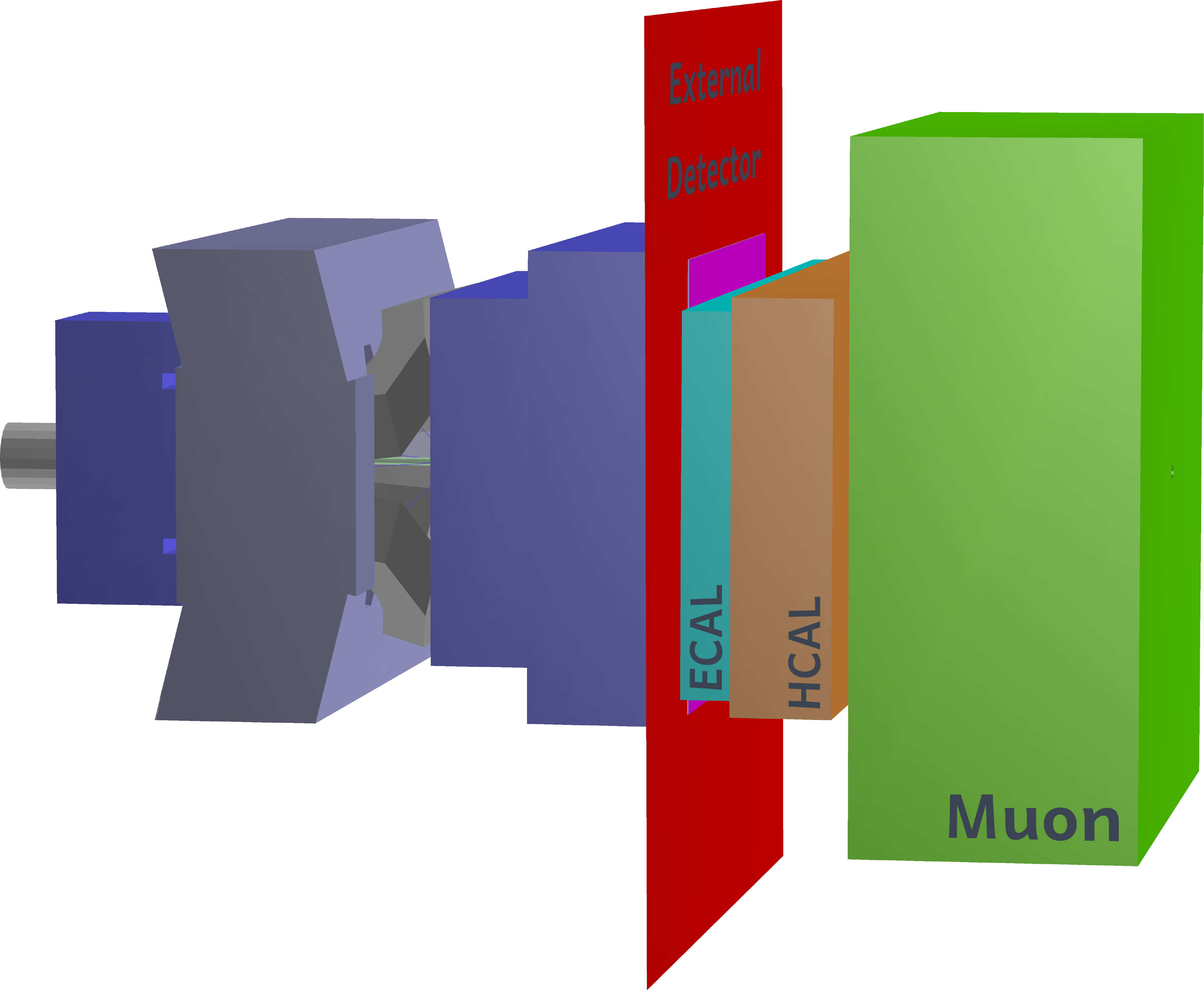}
    \caption{\lhcb upgrade  geometry~\cite{FastSimInterface} as seen by the \geant toolkit with an example of a plane-like detector (red, thin plane), introduced by the \externaldetector package. When used as a collector of particle information, it may provide the source of training information about incident particles for all the sub-detectors placed downstream from it along the beamline: \ecal (cyan box), \hcal (orange box), or muon system (green box). }
    \label{fig:upgrade_view_with_collector}
\end{figure}

\begin{figure}
    \centering
    \includegraphics[width=.93\textwidth]{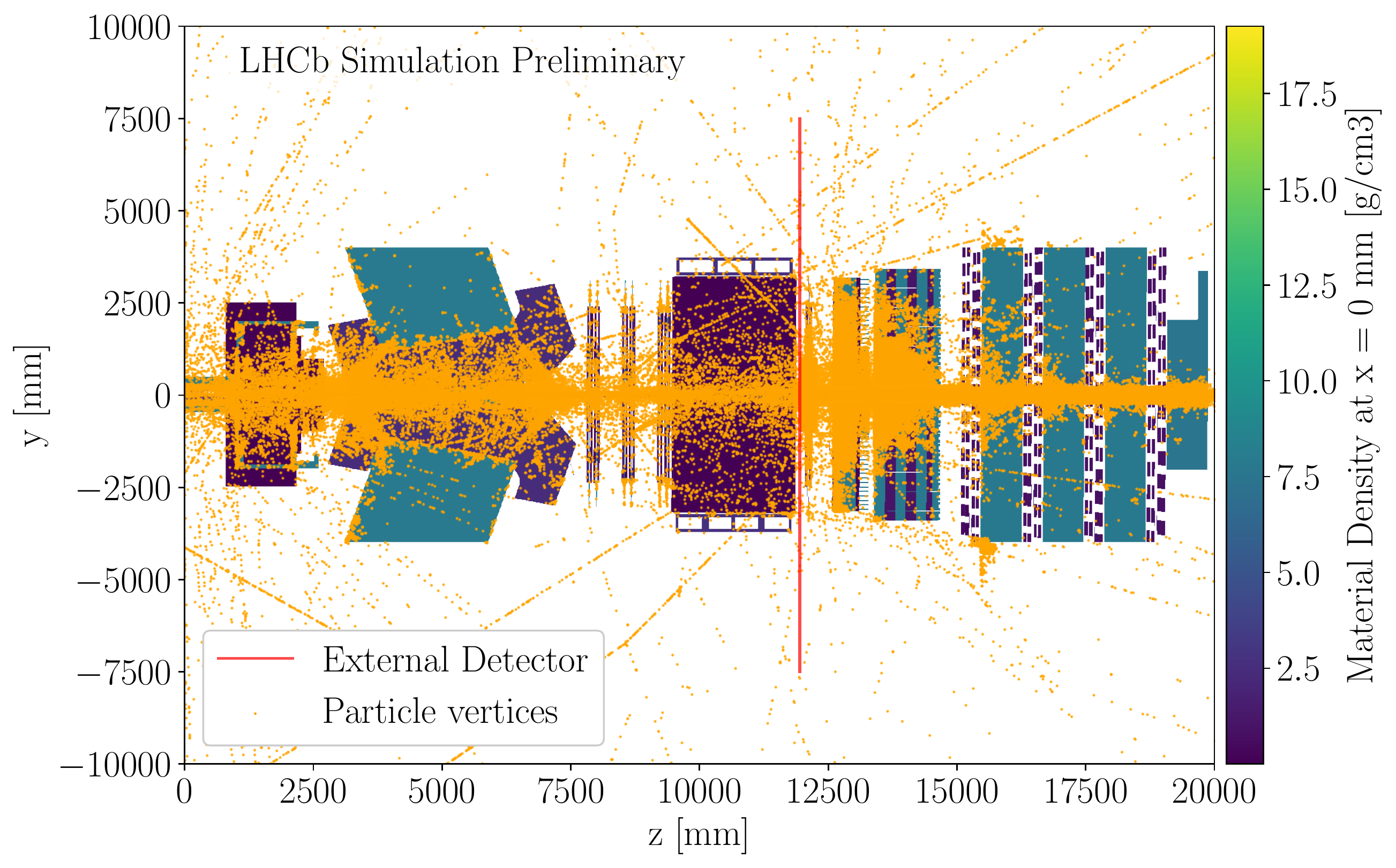}
\caption{Particles generated~\cite{FastSimInterface} using the current \gauss framework in the simulation of a minimum bias event with the beam conditions as foreseen in the Run 3 data-taking period and the upgrade geometry. An external plane-like detector that collects information about traversing particles is depicted with the red color.}
\label{fig:lhcb_upgrade}
\end{figure}

 Information about the simulated objects can be easily obtained by introducing a new sensitive sub-detector that would register hits of an abstract type with all the information needed to train the fast simulation model. A few difficulties may be encountered with this approach though. Since these detectors will only be used in just a few, specific studies, setting them up should be configurable on the fly without having to introduce them in the existing detector description, hosted in a database in an xml format. This functionality is provided by a new package in \gaussino called \externaldetector, that allows for sub-detectors of any shape to be inserted. An example of the external plane-like detector, as seen by  \geant, is illustrated in Figure \ref{fig:upgrade_view_with_collector}. A side view of the same setup is presented in Figure \ref{fig:lhcb_upgrade}, together with simulated particles information.
 
 The user can also choose what kind of factories should be attached to an external detector. In principle, the external detector can become a sensitive detector (i.e. it will be able to register hits) or the user can add a monitoring tool that will be launched at the end of each event in order to verify the integrity of the collected data. \mccollector provides a set of abstract sensitive detector factories that are easier to configure than those used in the standard simulation. Finally, it might be the case that the external detector will overlap with other existing volumes in the geometry. \parallelgeometry exploits an abstract concept introduced by \geant that allows for having multiple geometries in parallel, each of them performing the particle transport without interference from objects defined in other geometries. 
 
\begin{figure}
\begin{subfigure}[h]{\textwidth}
    \centering
    \includegraphics[width=.9\textwidth]{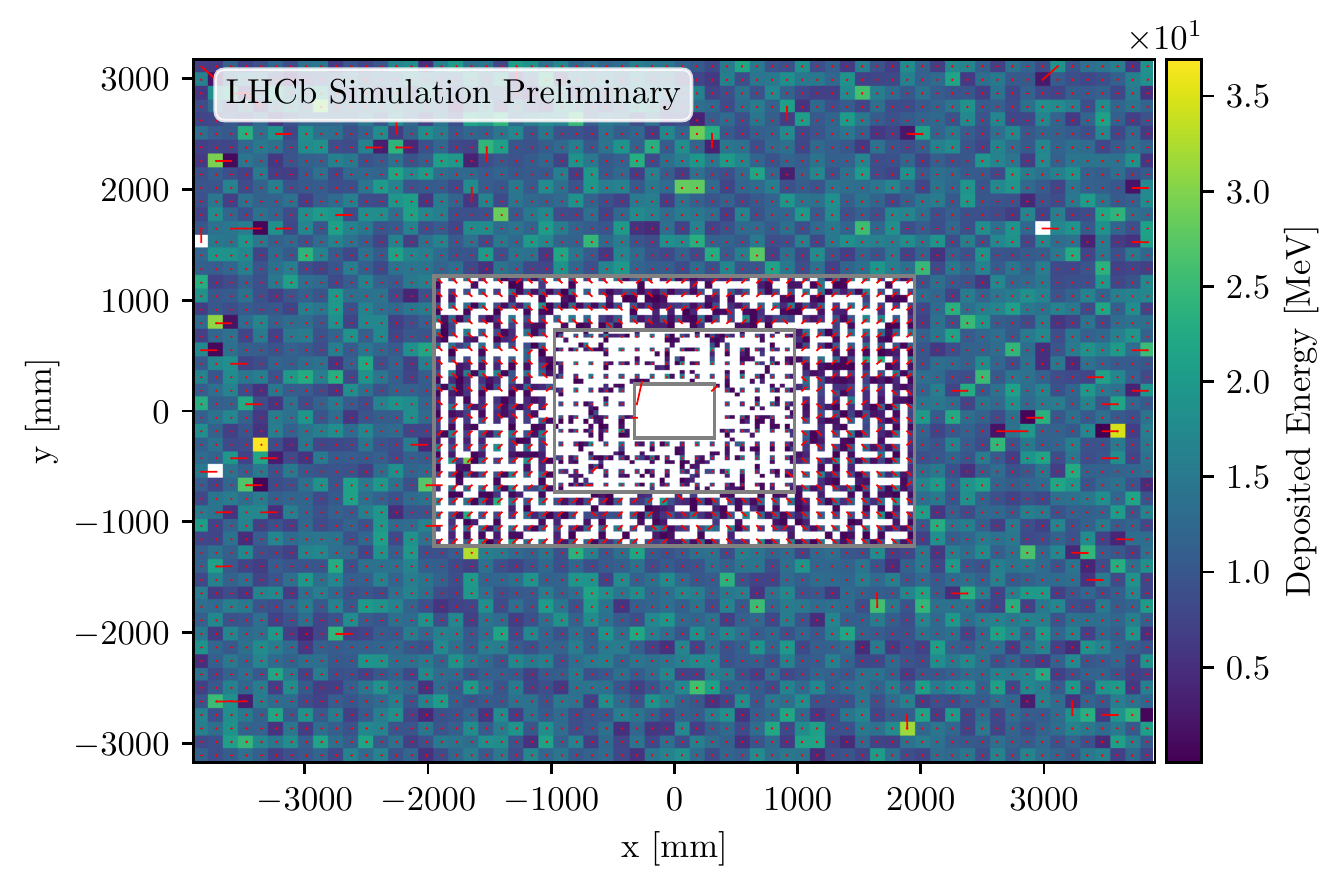}
    \caption{Particle gun with a grid of 3328 evenly-spaced 100 \mev photons.}
    \label{fig:training_example_pgun}
\end{subfigure}

\begin{subfigure}[h]{\textwidth}
    \centering
    \includegraphics[width=.9\textwidth]{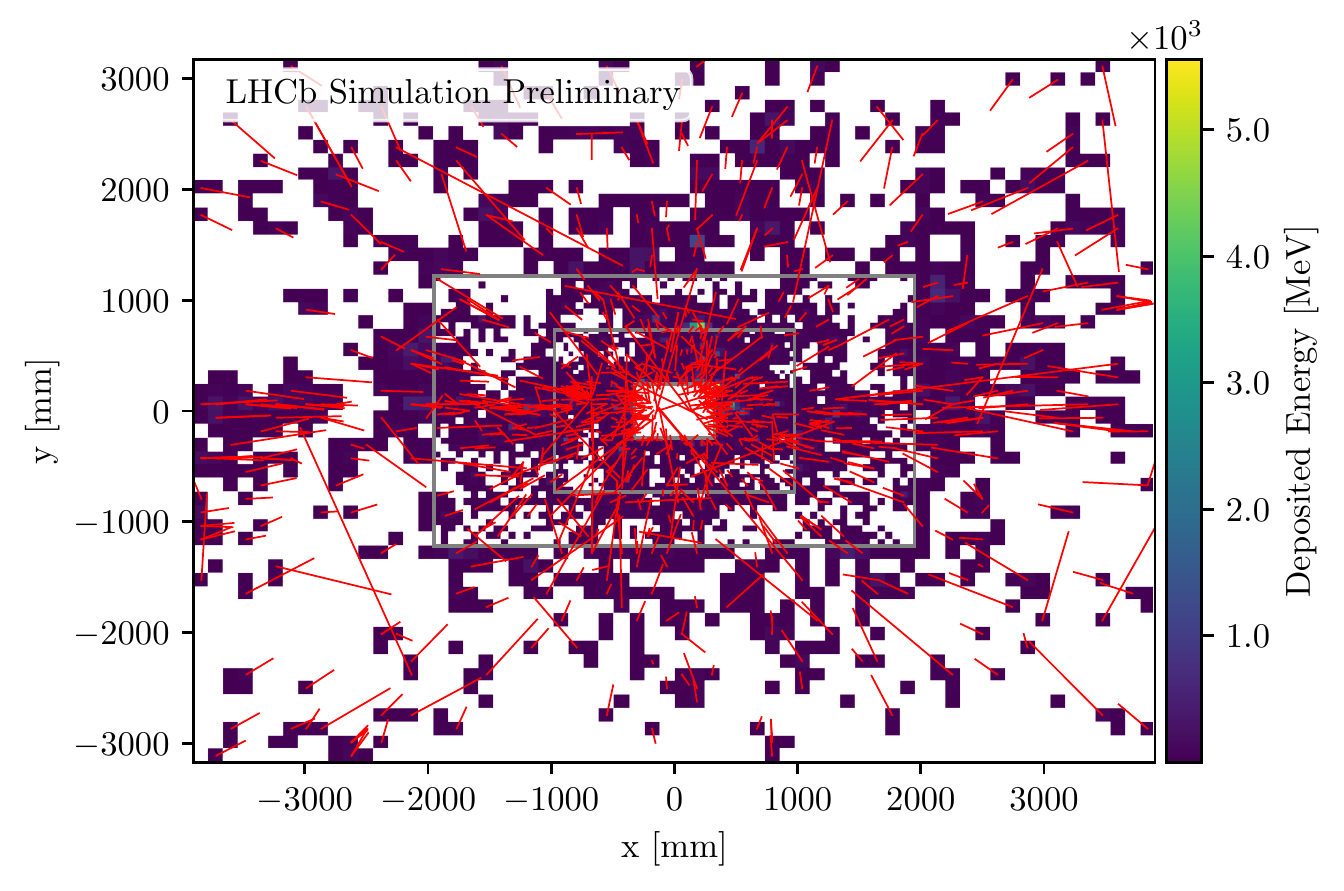}
    \caption{Minimum bias event with the beam conditions as foreseen in the Run 3 data-taking period and the upgrade geometry.}
    \label{fig:training_example_minbias}
\end{subfigure}
\caption{Visualization~\cite{FastSimInterface} of the training dataset produced by placing a collector plane in front of the electromagnetic calorimeter. Each of the images represent the ECAL energy deposits (hits) projected onto an $xy$-plane. The main role of the collector plane is to gather the positions of all particles intercepting the front face of ECAL. The particle positions are then linked (red lines) with the energetic centers of the showers generated by these particles. }
\label{fig:training_examples}
\end{figure}

The packages provide a generic way of producing the training datasets for fast simulation models. Two simple examples are presented in this paper in order to show how they can be used for the models under development for \ecal (e.g. point library \cite{PointLib} and GANs \cite{GANs}). Visual representations of the training datasets, produced by placing a collector plane in front of \ecal, are illustrated in Figure \ref{fig:training_examples}. In Figure \ref{fig:training_example_pgun}, a grid of 3328 evenly-spaced photons, similar to that used as an example when testing the performance of the interface, is shown. In Figure \ref{fig:training_example_minbias}, an input for the fast simulation studies requiring minimum bias events is presented.

This new interface will be exploited to provide training samples and integrate fast simulation models for given sub-detectors in \gaussino.
 
\section{Conclusions}

Monte Carlo simulation is a key component to model the effects of detector acceptance and imposed selection requirements. In order to address the computing constraints of the future upgrades of the \lhcb detector, the \lhcb simulation software \gauss has been redesigned to provide support for inter-event parallelism and integrate fast simulation models in \geant. In this paper, a \fastsimulation interface in \gaussino, the new experiment independent core simulation framework developed in \lhcb, was presented alongside with performance tests that show how the multi-threaded infrastructure of the fast simulation interface has a negligible contribution to the execution time. A general way of producing the training datasets for the fast simulation models using the \lhcb setup was also discussed. This new functionality will make it possible to integrate in the \gaussino framework the fast simulation models that are currently being developed in the \lhcb experiment as well as future ones.

\bibliographystyle{cai}
\bibliography{bib/LHCb-PAPER,bib/LHCb-DP,bib/LHCb-TDR,bib/standard,bib/custom}









\newpage
\appendix
\setcounter{secnumdepth}{0}
\section{Appendix}
\label{section:appendix}

\begin{figure}
\begin{subfigure}[h]{\textwidth}
    \centering
    \includegraphics[width=.85\textwidth]{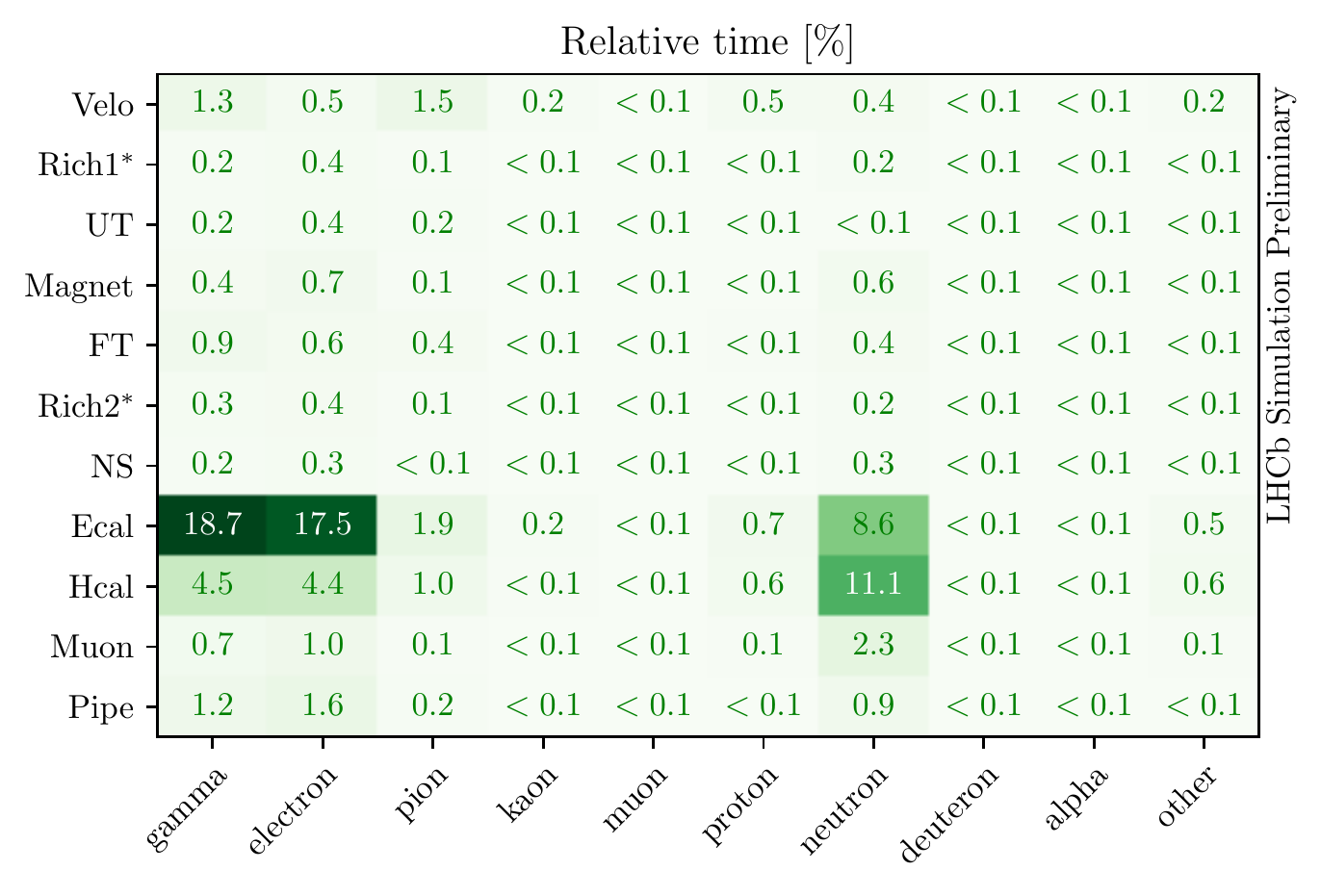}
    \caption{Relative time spent by each particle in a sub-detector measured with respect to the total time spent on the simulation.}
\end{subfigure}

\begin{subfigure}[h]{\textwidth}
    \centering
    \includegraphics[width=.85\textwidth]{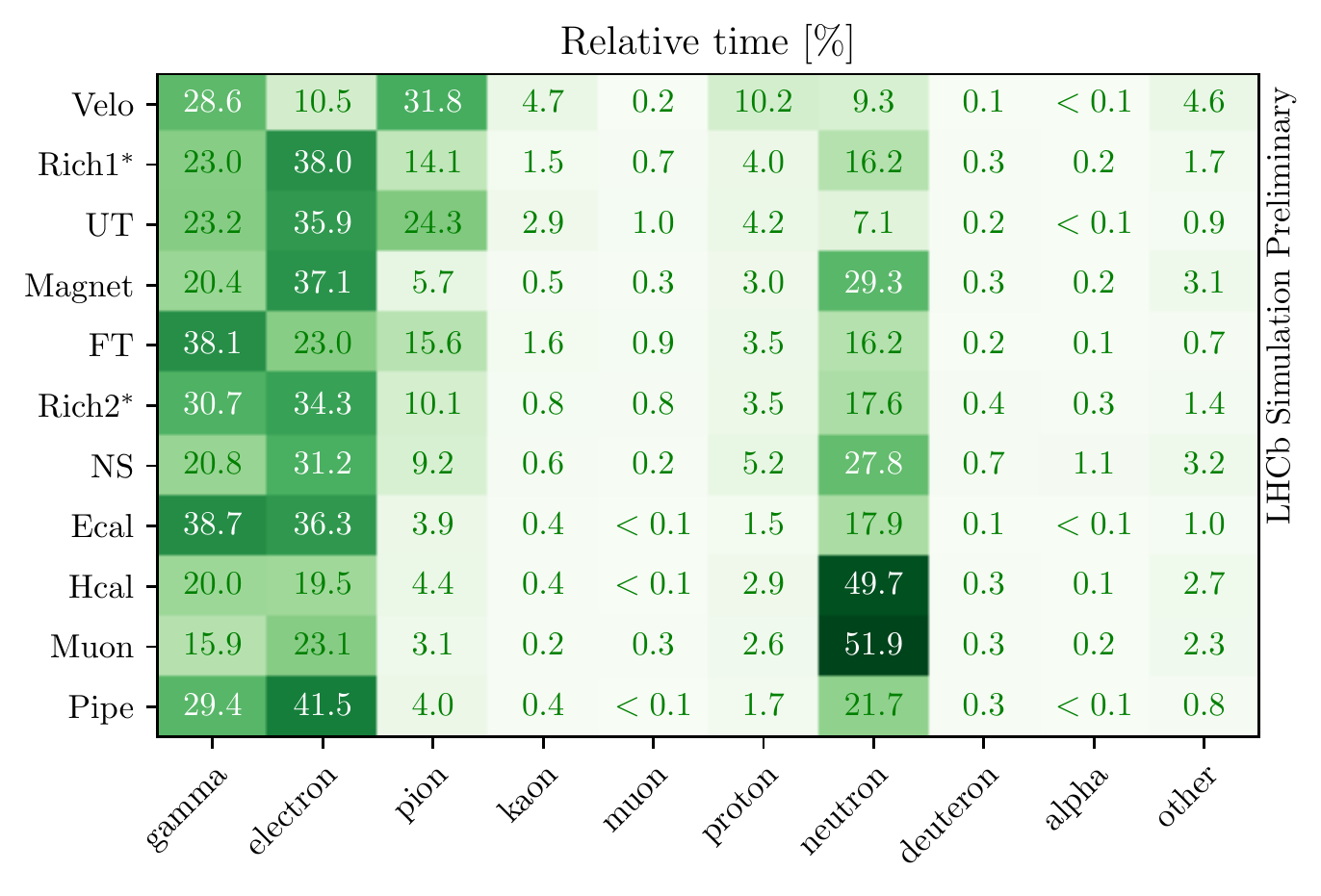}
    \caption{Relative time by each particle in a sub-detector measured with respect to the time spent in that sub-detector.}
\end{subfigure}
\caption{Performance~\cite{FastSimInterface} of the current \gauss framework when simulating 100 minimum bias events with the beam conditions as foreseen in the Run 3 data-taking period and the upgrade geometry.}
\label{fig:reltive_time_upgrade}
\end{figure}
\newpage
\begin{figure}
\begin{subfigure}[h]{\textwidth}
    \centering
    \includegraphics[width=.85\textwidth]{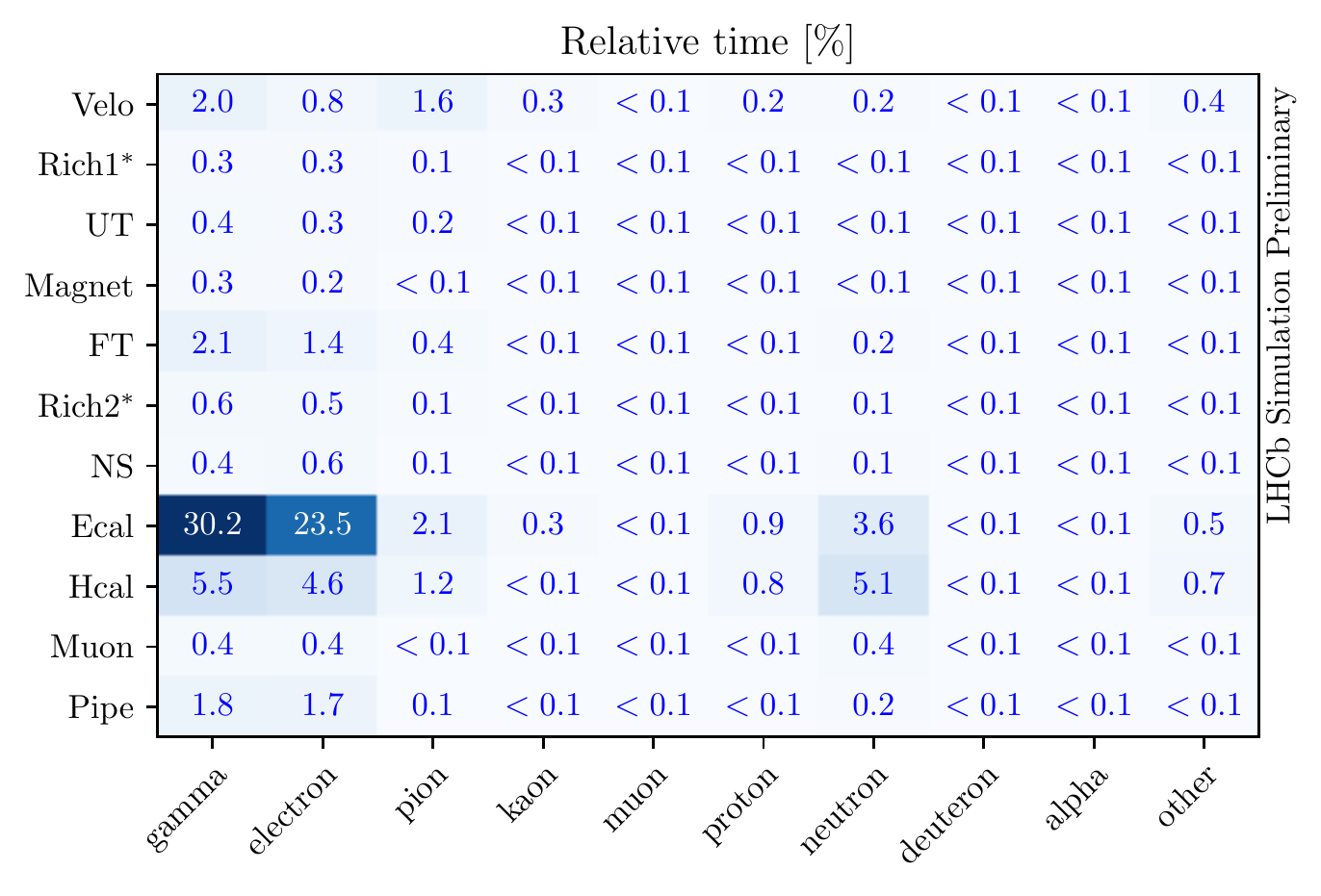}
    \caption{Relative time by each particle in a sub-detector measured with respect to the total time spent on the simulation.}
    
\end{subfigure}

\begin{subfigure}[h]{\textwidth}
    \centering
    \includegraphics[width=.85\textwidth]{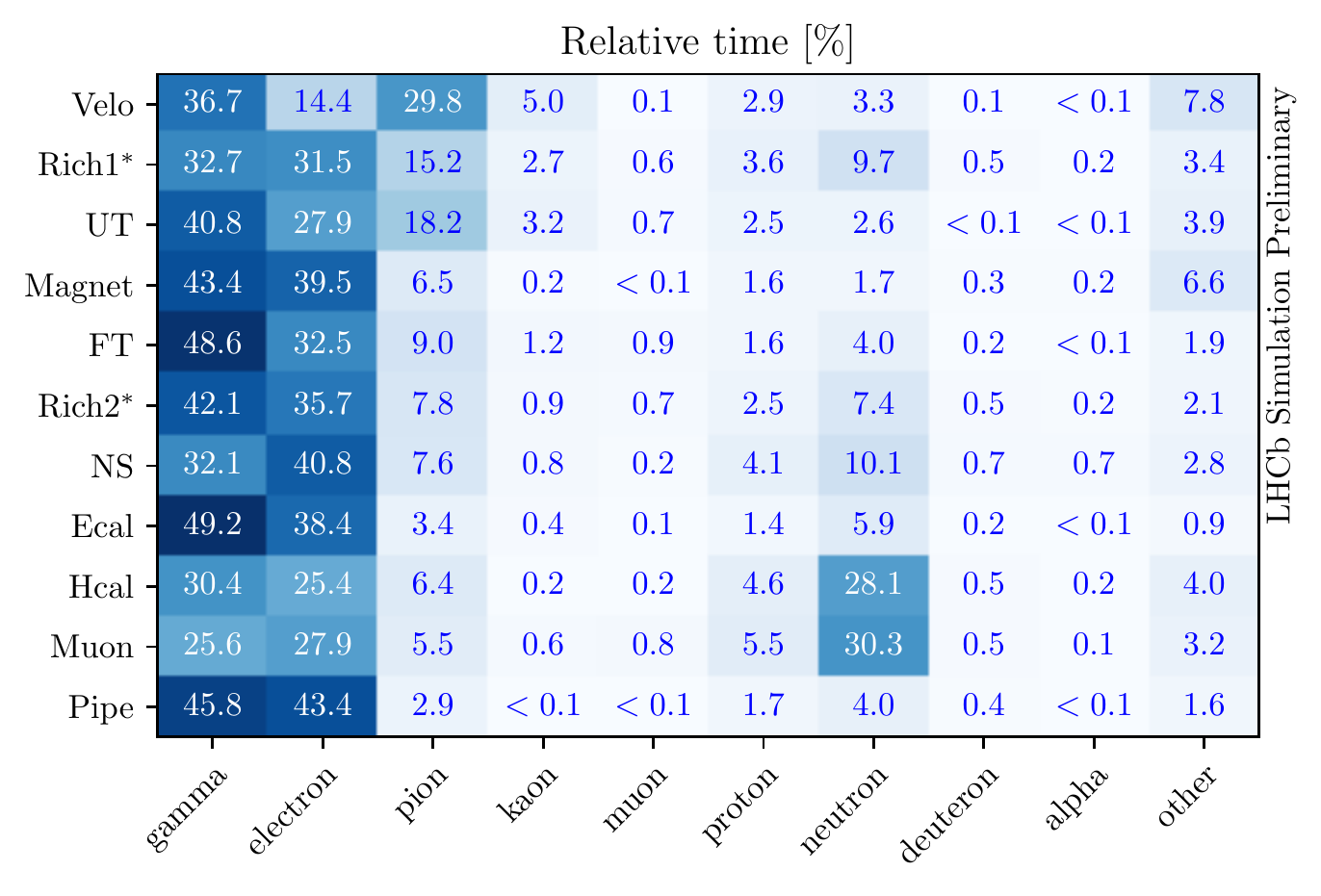}
    \caption{Relative time by each particle in a sub-detector measured with respect to the time spent in that sub-detector.}
    
\end{subfigure}
\caption{Performance~\cite{FastSimInterface} of the new multi-threaded (1 thread) \gaussino-based framework when simulating 100 minimum bias events with the beam conditions as foreseen in the Run 3 data-taking period and the upgrade geometry.}
\label{fig:reltive_time_2016}
\end{figure}

\label{lastpage}
\end{document}